\documentclass[11pt]{article}
\usepackage{amsmath,amsfonts}
\usepackage{pseudocode}

\newcommand\skipthis[1]{}

\begin{document}

\title{How inefficient can a sort algorithm be?}

\author{Miguel A. Lerma}

\maketitle

\begin{abstract}
Here
we find large lower bounds for a certain family of algorithms, and
prove that such bounds are limited only by natural computability
arguments.
\end{abstract}

\section{Introduction}

Here we study algorithms intended to sort a list of $n$ integers.  It
is well known that optimal sort algorithms such as \emph{mergesort}
have a run time $\Theta(n\log{n})$ (see \cite{knuth:taocp3}).
\emph{Bublesort}, with a worst-case run time of $\Theta(n^2)$, is
considered ``inefficient''.  But, are there any sort algorithms that
perform even worse?

This paper is inspired on a discussion found in Internet about
inefficient sort algorithms.  The summary of such discussion can be
found (at the time of this writing) in the following page:
\begin{quotation}
\texttt{http://home.tiac.net/\lower 3pt \hbox{${}^\sim$}cri\_d/cri/2001/badsort.html}
\end{quotation}

That discussion contains details on how to design sort algorithms with
larger than quadratic run time.  The record holder for such kind
of inefficient algorithm among the ones mentioned in that page is
called \emph{EvilSort}, with a run time $\Omega{((n^2)!)}$.
Here we show how to break that record and produce basically boundless
inefficient sort algorithms.

\section{A hierarchy of inefficient sort algorithms}\label{hierarchy}

Before we start stepping up the slope of inefficiency we want to make
sure that we don't do it in a trivial way, such as inserting useless
loops just with the purpose of ``wasting'' time by adding delays in an
artificial way.  The sort algorithms described here will always
contain only steps directed to the final goal of obtaining a sorted
list of elements.

The basic task of our algorithms will be to sort a list of integers
$L=[a_1,a_2,\dots,a_n]$ in increasing order.  The size of the input
will be given by the number $n$ of elements in the list, and time will
be measured by the number of integer comparisons performed.

A particularly inefficient way to sort the given list of integers
consists of generating a random permutation of it and check if such
permutation contains the elements correctly sorted.  That is the so
called \emph{bogosort} algorithm (see e.g \cite{gruber:stsw}),
performing asymptotically $(e-1)n!$ integer comparisons and $(n-1)
\cdot n!$ swaps in average.  This kind of algorithm however has
several problems. First, it requires a random generator.  Then, the
best case run time is very low, just $n-1$ integer comparisons and
no swaps if the given list is already sorted.  Finally, the worst case
run time is unbounded.

A variation of \emph{bogosort} that eliminates randomness consists of
generating all $n!$ permutations of the given list and then search for
the one that contains the elements correctly sorted.  This keeps the
average run time in $\Omega(n!)$ integer comparisons, but still
produces a low $n-1$ number of comparison in the best case.

In order to keep the best case run time high, we will change the
strategy to find the correctly sorted permutation.  Instead of
performing a linear search on the list of permutations, we will
\emph{sort} all $n!$ permutations in lexicographical order, an return
the first one of them.  For instance, if the given list is
$L=[2,3,1]$, we generate a list of lists consisting of all possible
permutations of the given list:
\[
P = [[2,3,1], [2,1,3], [3,1,2], [3,2,1], [1,2,3], [1,3,2]]
\]
and then sort them in lexicographical order:
\[
P_{\text{sorted}} = [[1,2,3], [1,3,2], [2,1,3], [2,3,1], [3,1,2], [3,2,1]] \,.
\]
The first element of this list of integer lists is the sorted integer
list $[1,2,3]$. 

The sorting of the list of integer lists can be performed with any
standard algorithm such as \emph{bublesort}, which runs in
$\Theta(n^2)$ time.  The lexicographical order of integer lists is
defined so that $L_1 <_{\text{lex}} L_2$ precisely when the first
index $k\in [1,\dots,n]$ for which they differ verifies $L_1[ k] <
L_2[ k]$.  So, comparing two integers lists requires at least one
integer comparison, and the total time (number of integer comparisons)
required to sort $n!$ permutations of $n$ elements in lexicographical
order using bublesort will be $\Omega((n!)^2)$.

That is still less than the run time of the EvilSort algorithm
mentioned above, but soon we will see how to do better---I mean,
worse.

One obvious way consists of replacing bublesort with another instance
of the algorithm just described, i.e., instead of using bublesort to
sort the $n!$ permutations of the original list of $n$ integers,
generate the $(n!)!$ permutations of the list of $n!$  permutations,
and then sort lexicographically the list of permutations of
permutations.  The first element will be a list of permutations of
integers lists, and the first element of it will be the original list
of $n$ integers sorted in increasing order.  The number of integer
comparisons performed will be now $\Omega(((n!)!)^2)$.

This finally breaks the record hold by EvilSort, but we want go
further, break our own record, and in fact any record ever set by
anybody in the past or in the future.  To do so we can repeat what we
just did, i.e., replace the final application of bublesort with an
instance of the latest version of the kind of algorithm described
here, so that the run time will keep growing to
$\Omega((((n!)!)!)^2)$, $\Omega(((((n!)!)!)!)^2)$, and so on,
but how far can we go?

In the next section we will develop these ideas in a more precise way,
and also will look at what the limit of this strategy might be.  In
particular we will answer the following question: given any (rapidly)
increasing computable function $f:\mathbb{N}\to \mathbb{N}$, is there
a sort algorithm with run time $\Omega(f(n))$?

\section{Worstsort: the final solution}

As stated, our algorithm will take as its input a list \texttt{L} with $n$
integer elements and return the same list with its elements sorted in
increasing order.  In the intermediate steps we will be handling
general lists whose elements can be of any type, in particular the
elements of a list can also be lists.  

The following is assumed about lists:

\begin{enumerate}

\item The number of elements of a list \texttt{L} is available
and represented as \texttt{length(L)}.

\item Elements are indexed with an index that runs from 1 to
\texttt{length(L)}.

\item It is possible to access/retrieve/modify the element at a
particular index without affecting any other elements.  In particular
it is possible to swap two elements of a list.

\item It is possible to insert an element at a particular index. The
indices of higher elements at that are increased by 1.

\item It is possible to remove an element at a particular index. The
indices of higher elements at that are decreased by 1.

\item It is possible to append two lists.  Here we represent \texttt{L1
+ L2} = result of appending list \texttt{L1} and \texttt{L2}, e.g.
$[a,b,c] + [d,e] = [a,b,c,d,e]$.

\end{enumerate}

The usual assignment operator '\texttt{:=}' between lists makes the
list in the left hand side identical to the list on the right hand
side, i.e., \texttt{L1 := L2} makes \texttt{L1} into another name for
list \texttt{L2}.  If after the assignment list \texttt{L2} is
modified, list \texttt{L1} is also modified because they in fact
represent the same list.

We can make a copy of a list in such as way that the original list and
its copy have the same elements, but remain different lists, so that
changes in the copy do not affect the original list.  The following is
an implementation of a list copy function (using Pascal-like
pseudocode):

\begin{pseudocode}
\PROCEDURE copy(A,B)
\FOR i := 1 \TO length(A)
\DO
\STATE B[i] := A[i]
\OD
\ENDPROCEDURE
\end{pseudocode}

The length and indexing of \texttt{B} are adjusted to fit those of
\texttt{A}.

Variables are supposed to be local to the procedure where they occur,
and created as needed if they do not exist. The types of variables
will be 'integer', 'list of integers', 'list of list of integers', and
so on.  The type of a variable is determined by context.  Integer
arguments are passed by value, and lists are passed by reference.

Since the algorithms to be precisely defined here will require not
only integer comparison, but also lexicographical comparisons of list
of integers, of lists of list of integers, etc., we need a function
\texttt{lt} that is able to perform that operation to any level.  The
following code fulfills this requirement:

\begin{pseudocode}
\PROCEDURE lt(A,B) \COMMENT is A less than B?
\IF type(A) = integer \THEN \COMMENT the arguments are integers
\RETURN{A<B} \COMMENT return integer comparison
\ELSE
\NLSTATE \COMMENT otherwise the arguments are lists,
\NLSTATE \COMMENT perform lexicographic comparison
\FOR k := 1 \TO length(A)
\DO
\IF lt(A[k],B[k]) \THEN
\RETURN{\TRUE} \COMMENT A[k] < B[k], hence A < B
\ELSIF lt(B[k],A[k]) \THEN
\RETURN{\FALSE} \COMMENT A[k] > B[k], hence A > B
\ELSE
\NLSTATE \COMMENT otherwise A[k] = B[k], keep going
\FI
\OD
\RETURN{\FALSE} \COMMENT all elements are equal, hence A = B
\FI
\ENDPROCEDURE lt
\end{pseudocode}

The following is the version of \texttt{bublesort} that we will be
using here. The algorithm modifies the original list \texttt{L}, and
performs $\Theta(n^2)$ '\texttt{lt}' comparisons.

\begin{pseudocode}
\PROCEDURE bublesort(L)
\FOR  i:=2 to length(L)
\DO
\FOR  j:=1 to length(L)-i+1
\DO
\IF lt(L[j+1],L[j]) \THEN
\STATE swap(L[j],L[j+1])
\FI
\OD
\OD
\ENDPROCEDURE bublesort
\end{pseudocode}

The procedure \texttt{permutations} takes a list as its argument and
returns a list of lists with all permutations of the elements of the
original list.  The following code is one among many possible ways of
generating all the permutations of a list \texttt{L}:

\begin{pseudocode}
\PROCEDURE permutations(L)
\IF length(L) =< 1
\THEN 
\NLSTATE \COMMENT in this case there is only one permutation
\STATE copy(L,L0) \COMMENT this is to preserve original list
\RETURN{[L0]} \COMMENT return the only permutation
\ELSE 
\STATE P := [] \COMMENT the list of permutations is initially empty
\FOR i:=1 \TO length(L)
\DO
\STATE copy(L,L1) \COMMENT make copy of original list
\STATE remove(i,L1) \COMMENT remove i-th element from the copy
\STATE P0 := permutations(L1) \COMMENT generate its permutations
\NLSTATE \COMMENT put removed element at the beginning
\NLSTATE \COMMENT of each permutation of L1 and add the
\NLSTATE \COMMENT result to the list of permutations
\FOR j:=1 \TO length(P0)
\DO
\STATE P := P + [[L[i]] + P0[j]]
\OD
\OD
\RETURN{P}
\FI
\ENDPROCEDURE permutations
\end{pseudocode}

The following is the code for the multilevel version of the sort
algorithm described in section~\ref{hierarchy}:

\begin{pseudocode}
\PROCEDURE multilevelsort(L,k)
\IF k = 0 \THEN \COMMENT last level, just perform bublesort
\STATE bublesort(L)
\ELSE
\STATE P := permutations(L) \COMMENT generate permutations
\STATE multilevelsort(P,k-1) \COMMENT sort them lexicographically
\STATE copy(P[1],L) \COMMENT copy first element into L
\FI
\ENDPROCEDURE multilevelsort
\end{pseudocode}

For $k=0$, \texttt{multilevelsort} performs just bublesort on the
given list of elements, run time $\Omega(n^2)$.  For $k>0$,
\texttt{multilevelsort} performs $k$ recursive self-calls before using
bublesort. Its run time is $\Omega(((\cdots(n!)\cdots!)!)^2)$, with
$k$ nested factorials.  Using the \emph{multifactorial} notation
$n!^{(k)}=$ take the factorial of $n$ $k$ times, then the lower bound
for the run time of \texttt{multilevelsort}
will~be~$\Omega((n!^{(k)})^2)$.

We finally answer the question of how inefficient a sort algorithm can
be.  To do so we define the following sort algorithm, that takes a
list of integers $L$, and an increasing computable function
$f:\mathbb{N}\to \mathbb{N}$ as its arguments:

\begin{pseudocode}
\PROCEDURE worstsort(L,f)
\STATE multilevelsort(L,f(length(L)))
\ENDPROCEDURE worstsort
\end{pseudocode}

The run time for this algorithm is now
$\Omega((n!^{(f(n))})^2) \geq \Omega(f(n))$, showing that a
sort algorithm can be made as inefficient as we wish, with its run
time growing at least as fast as any given fix computable function.
Since \text{worstsort} is itself computable, the growth rate of its
run time will still be asymptotically bounded above by rapidly
growing uncomputable functions such as a \emph{busy beaver} (which is
known to grow faster than any computable function---see
\cite{rado:oncf}).  But given any fix rapidly growing
\emph{computable} function, we can make the run time of
\text{worstsort} grow faster just by feeding that function as its
second argument.

\section{Conclusion}

We have shown that there is no computable limit to the inefficiency of
a sort algorithm, even when respecting the rule of not using useless
loops and delays unrelated to the sorting task.  The run time of
such algorithm can growth at least as fast as any given fix computable
function.

\newpage

\bibliographystyle{plain} \bibliography{comp}

\end{document}